\documentclass{ws-ijmpd}

\usepackage{lscape}
\newcommand{\be}{\begin{equation}}
\newcommand{\ee}{\end{equation}}
\newcommand{\bea}{\begin{eqnarray}}
\newcommand{\eea}{\end{eqnarray}}

\begin{document}

\markboth{Melia and L\'opez-Corredoira}
{AP Test with BAO}

%
\catchline{}{}{}{}{}
%

\title{Alcock-Paczy\'nski Test with Model-independent BAO Data}

\author{F. Melia$^1$ and M. L\'opez-Corredoira$^{2,3}$}

\address{$^1$ Department of Physics, The Applied Math Program, and Department of Astronomy,
The University of Arizona, AZ 85721, USA;\\
$^2$ Instituto de Astrof\'\i sica de Canarias,\\ 
E-38205 La Laguna, Tenerife, Spain\\
$^3$ Departamento de Astrof\'\i sica, Universidad de La Laguna,\\
E-38206 La Laguna, Tenerife, Spain}

\maketitle

\begin{history}
\received{Day Month Year}
\revised{Day Month Year}
\comby{Managing Editor}
\end{history}

\begin{abstract}
Cosmological tests based on the statistical analysis of galaxy distributions usually 
depend on source evolution. An exception is the Alcock-Paczy\'nski (AP) test, which 
is based on the changing ratio of angular to spatial/redshift size of (presumed) 
spherically-symmetric source distributions with distance. Intrinsic redshift 
distortions due to gravitational effects may also have an influence, but they can 
now be overcome with the inclusion of a sharp feature, such as the Baryonic Acoustic 
Oscillation (BAO) peak. Redshift distortions affect the amplitude of the peak, but 
impact its position only negligibly. As we shall show here, the use of this 
diagnostic, with new BAO peaks from SDSS-III/BOSS at average redshifts $\langle 
z\rangle=0.38$, $0.61$ and $2.34$, disfavors the current concordance ($\Lambda$CDM) 
model at $2.3\sigma$. Within the context of expanding Friedmann-Robertson-Walker 
(FRW) cosmologies, these data instead favor the zero active mass equation-of-state, 
$\rho+3p=0$, where $\rho$ and $p$ are, respectively, the total density and pressure 
of the cosmic fluid,  the basis for the $R_{\rm h}=ct$ universe. 
\end{abstract}



\section{Introduction}
A highly desirable goal in cosmology is the acquisition of model-independent data that can
be used to test theoretical models and to optimize their parameters. There are two main
types of cosmological observations that shed light on the geometry of the Universe:
a measurement of the fluctuations in the Cosmic Microwave Background Radiation (CMB) and
the analysis of large-scale structure via the inferred distribution of galaxies.
The quality and quantity of both kinds
of data have progressed significantly in recent decades. CMB anisotropies provide the
most evident support for the concordance ($\Lambda$CDM) model, but one should find
an independent confirmation of this theory and its parameters because CMB anisotropies
may be generated/modified by mechanisms other than those in the standard picture\cite{7,8,9,10}, 
and may also be contaminated by other effects\cite{11}. Cosmological
tests using surveys of galaxies have also been developed to provide information on the
geometry of the Universe: for instance, the Hubble diagram or the angular-size test.
However, both of these need some assumptions in order to provide cosmological information.

As of today, Hubble diagrams constructed from the apparent magnitude (taking into
account K-corrections) vs. distance or redshift of first-rank elliptical galaxies
in clusters require a strong evolution of galaxy luminosities to fit FRW cosmologies\cite{1}. 
For Type Ia Supernovae\cite{12} (SNIa) embedded
in these galaxies, or for gamma-ray bursts\cite{13}, one assumes zero evolution, and
the standard model fits the data, though some systematic effects may be affecting the
result, including a possible evolution in the metallicity of SNIa progenitors\cite{2}, possible 
internal extinctions\cite{14},
time variations of the grey-dust absorption of light from these supernovae in various
types of host galaxies\cite{15}, and observational selection effects.
The angular-size vs. redshift test has been developed by several authors\cite{3},
using sources seen at radio, near-infrared and visible wavelengths. All these applications
produce an angular size $\theta \sim z^{-1}$ up to $z\sim 3$, suggesting a strong
evolution in galactic radii. The fact that galaxies with the same luminosity apparently
were six times smaller at $z=3.2$ than at $z=0$ makes it
difficult to compare different models.\cite{4} The surface brightness (known as the `Tolman')
test also depends strongly on the assumption of galaxy evolution, so the results
of this test\cite{5,6} may vary hugely depending on one's interpretation.

An alternative cosmological test based on the Alcock \& Paczy\'nski (AP)\cite{16,17}
approach evaluates the ratio of observed angular size to radial/redshift size. 
This probe is based on the assumption that the distribution of sources is spherical,
regardless of distance, and large, so that their size can be easily measured over
a broad range of redshifts, preferably extending beyond $2-3$. Appropriate targets 
therefore include clusters of galaxies, rather than the individual galaxies 
themselves, like in some of the other tests described above. The main advantage 
of the AP test is that, regardless of whether or not galaxies may have evolved 
with redshift, it depends only on the geometry of the Universe. However, redshift
distortions induced by peculiar velocities\cite{18,19,20}, described in linear 
perturbation theory by the parameter $\beta$, can also have an influence. 
In the linear regime, the value of $\beta$ measured from redshift distortions 
corresponds to the solution of the linearized continuity equation
($\beta \delta +\nabla \vec{v}=0$, where $\delta $ is the relative overdensity, 
and $\vec{v}$ is the velocity field)\cite{20}. 

There is now a way to overcome 
this possible contamination---via the inclusion in the AP test of an effect with a 
sharp feature, such as the Baryon Acoustic Oscillation (BAO) peak, for which 
the degeneracy between redshift and geometric and gravitational distortions is 
almost completely broken\cite{21}. The reason is that the value of $\beta $ 
affects primarily the amplitude of the peak, but its impact on the position is
negligible. In this paper, we carry out the AP test using three Baryon Acoustic 
Oscillation (BAO) peak positions that significantly increase its precision over what 
has been achieved thus far. In so doing, we demonstrate that the concordance 
($\Lambda$CDM) model is disfavored by these new data. Instead, the AP test favors
a model with zero active mass, i.e., with the equation of state $\rho+3p=0$,
where $\rho$ and $p$ are, respectively, the total density and pressure of the
cosmic fluid\cite{22,23,24,25}.

\section{The Alcock-Paczy\'nski Test}
Given a spherically symmetric structure, or distribution of objects, with radius
\begin{equation}
s_\parallel=\Delta z\frac{d}{dz}d_{\rm com}(z)
\end{equation}
along the line of sight and a radius
\begin{equation}
s_\perp =\Delta \theta (1+z)^m d_{\rm A}(z)
\end{equation}
(where $m=1$ with expansion, while $m=0$ for a static universe) perpendicular
to the line of sight, the ratio
\begin{equation}
y\equiv \frac{\Delta z}{z\Delta \theta }\frac{s_\perp}{s_\parallel }
\end{equation}
depends only on the cosmological comoving distance, $d_{\rm com}(z)$, and the
angular-diameter distance, $d_{\rm A}(z)$, and is independent of any source
evolution.

Previous applications of the galaxy two-point correlation function
to measure a redshift-dependent scale for the determination of $y(z)$
were limited by the difficulty in disentangling the acoustic length
in redshift space from redshift distortions due to internal gravitational
effects\cite{17}. A serious drawback with this process
is that inevitably one had to either pre-assume a particular model, or
adopt prior parameter values, in order to estimate the level of
contamination. And the wide range of possible distortions (i.e., values
of $\beta$; see, e.g., Eqns.~4 and 5 in ref.~\cite{17}) for the
same correlation-function shape resulted in seriously large errors.

But this situation has changed dramatically in the past few years with 
the use of reconstruction techniques\cite{Eis07,Pad12} that enhance the 
quality of the galaxy two-point correlation function and the much more
precise determination of the Ly-$\alpha$ and quasar auto- and
cross-correlation functions, resulting in the measurement of BAO
peak positions to better than $\sim 4\%$ accuracy. In this paper,
we determine $y(z)$ using three BAO peak positions: 1) the
measurement of the BAO peak position in the anisotropic distribution
of SDSS-III/BOSS DR12 galaxies\cite{Ala16} at the two 
independent/non-overlapping bins with $\langle z\rangle=0.38$ and
$\langle z\rangle=0.61$, in which a technique of reconstruction
to improve the signal/noise ratio was applied. This technique affects the position 
of the BAO peak only negligibly, so the measured parameters are independent 
of any cosmological model. And 2) the self-correlation of the BAO peak in the
Ly-$\alpha$ forest in the SDSS-III/BOSS DR11 data\cite{27} at
$\langle z\rangle=2.34$, plus the cross-correlation of the BAO
peak of QSOs and the Ly-$\alpha$ forest in the same survey\cite{21}.

In all of these measurements, the template used for the correlation
function was drawn from the concordance model. However, the actual shape
of the BAO peak does not significantly affect the calculation of its centroid position,
both along the line-of-sight and in the direction perpendicular to it,
when its FWHM is very narrow. Any shape could be used, and the results
would be the same. The peak's narrowness strongly reduces the impact
of redshift distortions, which affect the peak's amplitude, but not so much its
location. Any modifications to this amplitude as a function of distance
may produce a second-order shift to the centroid, but always negligibly.
Reconstruction also significantly reduces the effects of redshift-space 
distortions at the BAO scale, isotropizing the correlation function\cite{Pad12}.
This conclusion is reflected, for instance, in the fact that, although the
errors for the parameter $\beta$ quoted in Table 2 of ref.~\cite{27}
are very large, the relative error bars for $d_A(z)$ and $H(z)$ are much
smaller. If the BAO peak measurements were sensitive to $\beta$, their error
bars would be much bigger. The redshift distortions produce systematic 
errors of order of 1-2\% in $d_A(z)$ and $H(z)$\cite{Tar10}, 
which are already included in the errors we use for the data, 
though some authors\cite{Sab16} give a slightly more pessimistic 
estimation of the error associated with the Hubble parameter. But
this situation has improved even more recently\cite{Alam16}, 
with the report of systematic uncertainties in $D_A/H$ that are even smaller:
$0.5\%$ for the AP parameter and $0.4\%$ for the distance scale---and 
these were considered conservative. This impressive precision amplifies
the probative power of these measurements, given that statistical errors
as small as $\sim 4\%$ are now achievable.

The angular-diameter distance $d_A(z)$ and Hubble constant $H(z)$ are related 
to the variable $y(z)$ via the expression
\begin{equation}
\label{y}
y(z)=\left(1+\frac{1}{z}\right)\frac{d_A(z)^*H(z)}{c}\;,
\end{equation}
where $d_A(z)^*$ is the measured angular-diameter distance assuming an expanding Universe,
that is, $d_A(z)^*=d_A(z)$ if there is expansion and $d_A(z)^*=\frac{d_A(z)}{(1+z)}$
for a static universe. Very importantly, the quantity $y(z)$ is independent
of the uncertain (co-moving) acoustic scale $r_s$, since $d_A(z)^*\propto r_s$,
while $H(z)\propto 1/r_s$, so the dependence on $r_s$ cancels out in the product.
(Sometimes an alternative definition of this ratio, $F_{AP}(z)\equiv z y(z)$, 
has been used in the literature, but these properties are the same for both working 
definitions.)

At low redshift, the values of the $F_{AP}(z)$ of ref.~\cite{Ala16} lead directly 
to $y(z=0.38)=1.079\pm 0.042$, $y(z=0.61)=1.248\pm 0.044$. At high redshift, we have
$c/H(z=2.34)r_s=9.15^{+0.20}_{-0.21}$ and $d_A(z=2.34)/r_s=10.93^{+0.35}_{-0.34}$  
(see Eqns.~22 and 23 in ref.~\cite{27}), plus the adoption of a correlation 
coefficient\cite{27} of -0.6 between $c/H$ and $d_A$, which lead to $y(z=2.34)=1.706\pm 
0.083$. The uncertainties in $y(z)$ are found by propagating the errors in $H(z)$ and 
$d_A(z)$ through Equation~(\ref{y}) including the covariance terms derived from the 
given correlation coefficients.

These three are the only BAO measurements currently available that
measure both $d_A(z)$ and $H(z)$ with small statistical errors.
Other reported values (e.g., ref.~\cite{26}) are older, less accurate versions 
of these same BOSS measurements. The available WiggleZ data\cite{Hin16} are not 
included because they are much less accurate than BOSS data: they have relative 
error bars of 20-30\% for $y(z)$. And we do not include clustering measurements 
of $d_A(z)$ and $H(z)$ based on the anisotropic two--point correlation function 
at shorter scales\cite{28,29,30,San14,Sam14,Beu14} because these are biased 
by the pre-assumption of a cosmological model, or the adoption of priors, used to 
characterize the internal redshift distortions (i.e., the parameter $\beta$).

\begin{figure}[t]
\centering
\includegraphics[width=13.3cm]{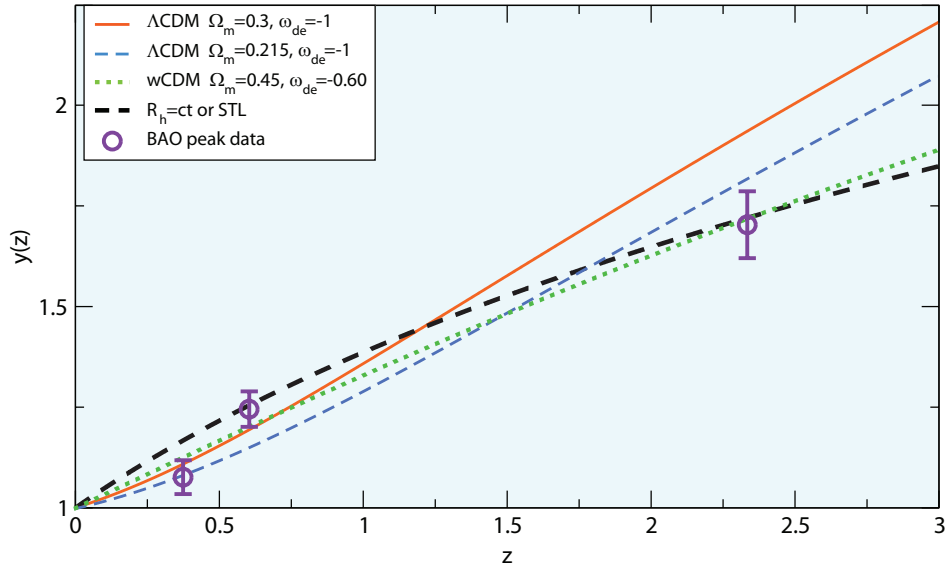}
\caption{Plot of $y(z)$ (the ratio of observed angular size to
radial/redshift size) versus $z$, inferred from the data using
the BAO peak measurements. This figure also shows the predictions
of the concordance ($\Lambda$CDM) model with fiducial parameter values
$\Omega_{\rm m}=0.3$ and $w_{\rm de}\equiv w_\Lambda=-1$,
a best fit using $\Omega_{\rm m}$ as a free parameter (but with a
fixed $w_\Lambda =-1$), and a best fit when both $\Omega_{\rm m}$
and $w_{\rm de}$ are free. The (heavy) dashed curve
represents the (expanding) $R_{\rm h}=ct$  and (static)
STL cosmologies, whose $y(z)$ functions are
(coincidentally) identical (see text).} 
\vspace{0.2cm}
\end{figure}

In figure~1, we compare various model predictions with these three BAO 
measurements of $y(z)$, following the conventional methodology of the
AP cosmological test\cite{16,17}. We here plot the function for the
wCDM model (the version of $\Lambda$CDM with a dark-energy equation of state
$w_{\rm de}\equiv p_{\rm de}/\rho_{\rm de}$ different from $w_\Lambda=-1$),
in addition to the predictions of the standard $\Lambda$CDM model itself with
fiducial parameter values $\Omega_{\rm m}=0.3$ and $w_{\rm de}\equiv w_\Lambda=-1$.
Additional comparisons are provided using $\Lambda$CDM with a non-concordance
value of $\Omega_{\rm m}$ as the sole free parameter, and also using fits with
both $\Omega_{\rm m}$ and $w _{\rm de} $ free. Table 1 summarizes
the quality of the fits plotted in figure~1. Additionally, we plot
$y(z)$ for the $R_{\rm h}=ct$ cosmology\cite{22,23,24,25}, in which
\begin{equation}
d_{\rm com}(z)=(1+z)d_A(z)=\frac{c}{H_0}\ln (1+z)\;,
\end{equation}
and has a $y(z)$ coincident with that of the Static/Tired Light (STL) model\cite{17},
though its representation of the Universe is based on the FRW metric with expansion,
while STL is static. This coincidence arises because for STL
$d_{\rm com}(z)=d_A(z)=\frac{c}{H_0}\ln (1+z)$, which corresponds to
a factor $(1+z)$ difference with the angular diameter distance $d_A(z)$ in
the $R_{\rm h}=ct$ universe. The latter also has a Hubble parameter $H(z)=H_0(1+z)$,
while $H$ is constant in STL. So the various factors of $(1+z)$ all cancel
in the formulation of $y(z)$ when calculated with $d_A^*(z)$ according to Equation~(\ref{y}).
The $R_{\rm h}=ct$ model represents a cosmology with zero
`active mass,' i.e., with the equation of state $\rho+3p=0$, where $\rho$ and $p$
are, respectively, the total density and pressure of the cosmic fluid. This model
has successfully passed all other cosmological tests applied to it thus 
far\cite{31,32,33,34,35,36,37,38}, though there remain some observations 
to be understood. 

\vskip 0.2in
\noindent {\footnotesize{\bf Table 1.} Results of the $\chi ^2$-test, using the 
$N=3$ points of the BAO peak data: the minimum reduced $\chi _{\rm red}^2
\equiv{\chi ^2}/{(N-\nu)}$, where $\nu$ is the number of free parameters; 
best-fit free parameters (if any); and associated probability of the models.}
\vskip -0.2in
\begin{table}
\begin{center}
{\footnotesize
\begin{tabular}{lcll}
&&& \\
\hline\hline
&&& \\
Model & $\;\;\chi _{\rm red,min}^2\;$ & Free parameters & Probability \\
&&& \\
\hline\hline
&&& \\
$\Lambda $CDM; $\Omega_{\rm m}=0.3$, $w_{\rm de} =-1$ & 3.25 & ---  & 0.0207 \\
$\Lambda $CDM; $\Omega_{\rm m}$ free; $w_{\rm de}=-1$  & 3.31 & $\Omega_{\rm m}=0.215^{+0.045}_{-0.040}$ & 0.0367 \\
wCDM; $\Omega_{\rm m}$, $w_{\rm de}$ free & 2.38 & $\Omega_{\rm m}=0.45^{+0.21}_{-0.19}$,
$w_{\rm de}=-0.60^{+0.30}_{-0.27}$  & 0.127 \\
$R_{\rm h}=ct$ (or STL) & 1.58 & ---  & 0.192 \\
&&& \\
\hline\hline
\end{tabular}
}
\end{center}
\end{table}

\section{Discussion}
\subsection{Model Comparison with the Alcock-Paczy\'nski Test}
In Table 1, we list the outcome of our model comparisons based on
the optimized fits to the data shown in figure~1, using a minimization
of the $\chi^2$ statistic and a calculation of the probability of each 
model being correct when the relative number of free parameters is taken 
into account. Each is an absolute probability, independent of the other 
models, and the confidence limits are extracted from the $\chi^2$-distribution. 
The $R_{\rm h}=ct$ universe (and, coincidentally, also the static 
STL cosmology) fits the data very well without any ad hoc optimization 
of free parameters. For the $\Lambda$CDM/wCDM cosmology, we see 
that the fiducial parameter values ($\Omega_{\rm m}=0.3$, $w_\Lambda =-1$) 
are excluded at a $97.93$\% C.L. (equivalent to 2.3$\sigma $), while even 
allowing $\Omega_{\rm m}$ to be free with a fixed $w_\Lambda =-1$ (best 
fit for $\Omega_{\rm m}=0.215^{+0.045}_{-0.040}$) is excluded at 
1.7$\sigma $. The only way to reconcile the data with the wCDM cosmology 
at an exclusion C.L. $<95$\% is to set $w_{\rm de}=-0.60^{+0.30}_{-0.27}$, 
higher than the fiducial value $w_{\rm de}=-1$, and $\Omega_{\rm m}=
0.45^{+0.21}_{-0.19}$. 

The confidence level contours for the standard model are shown in figure~2.
Note that the probabilities given in Table 1 are $P(\chi^2,N-\nu)$. However,
in figure~2 the probabilities of the fits are calculated as $P(\chi^2,N)$
since the parameters are here assumed to have prior values. That is, 
in figure~2 we are not taking into account the reduction of the probability 
due to a reduction in the number of degrees of freedom, since the parameters 
are not being optimized freely. It is done this way in order to compare the 
outcome using standard values ($\Omega_{\rm m}=0.3$, $w_{\rm de}=-1$) with 
those where the parameters are free. 

\begin{figure}[t]
\centering\includegraphics[width=4.0in]{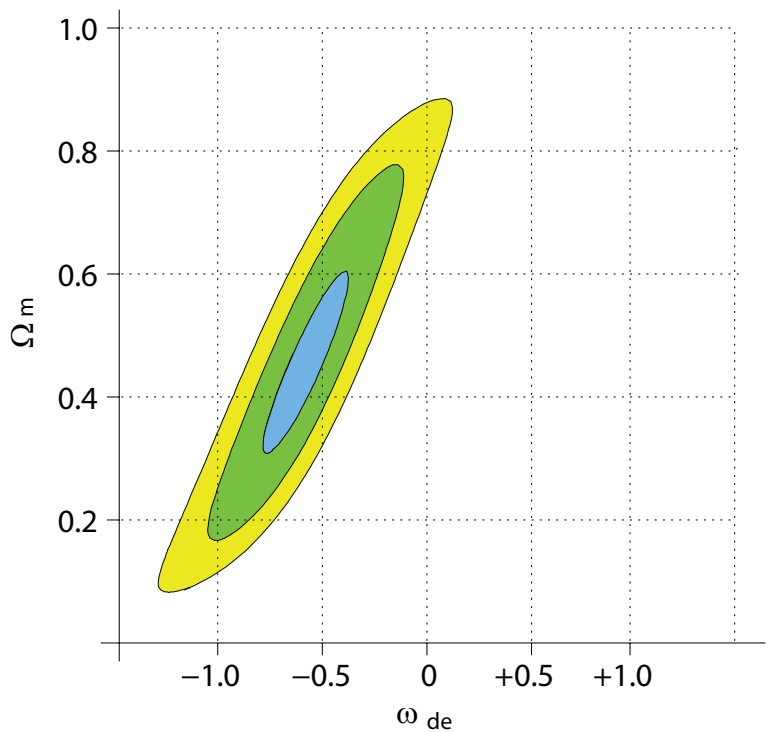}
\caption{Confidence level contours as a function of $\Omega _m$ and $w _{\rm de}$
for the $\Lambda$CDM/wCDM model. The three levels are equivalent to 1-3 $\sigma $ in a Gaussian
distribution (i.e., 68.3\% C.L., cyan; 95.4\% C.L., green; 99.73\% C.L., yellow). 
The concordance ($\Lambda$CDM) model corresponds to the optimized parameter values 
$\Omega _m=0.3$ and $w _{\rm de}=-1$, which are excluded at the 97.93\% C.L.}
\label{figure2}
\end{figure}

Our results reinforce the conclusions drawn in refs.~\cite{27,39,40}, who also 
remarked on the tension between the measurement
at $z=2.34$ and predictions of the standard model, based on various methods of analysis.
Abdalla et al.\cite{39} also pointed out that values of $w_{\rm de}>-1$ were necessary
to alleviate the tension for the datum $H(z=2.34)$, while Aubourg et al.\cite{40} realized
this tension, but did not explore such possible solutions because they fitted the
cosmological parameters using a combination of BAO+(Supernovae data and/or the CMB).
However, our results are based on the simultaneous analysis of the available 
model-independent BAO data (at $\langle z\rangle=0.38$, $z=0.61$ and $2.34$),
which more emphatically exclude the standard model, while favoring the $R_{\rm h}=ct$ 
universe. In this paper, we do not combine the AP test data with the CMB or 
Supernova data, because this step would dilute the information concerning the 
anomaly reflected in our results.

Our conclusions are stronger than the tension reported in 
refs.~\cite{27,39,40}, for the principal reason that the Alcock-Paczy\'nski 
test amplifies the deviation of the measurements relative
to the $\Lambda$CDM prediction. These authors reported a tension at the
$\gtrsim 2\;\sigma$ level. The AP test, however, compares the observed product
$d_A(z)\,H(z)$ with that of the model, and since both BAO measurements of $d_A(z)$
and $H(z)$ are higher than those expected in the standard model, the
AP discrepancy is larger than simply that of the quadrature sum of the
two individual discrepancies. And while Aubourg et al.\cite{40} did not
find any models that substantially improved the agreement between theory
and the Ly-$\alpha$ forest BAO measurement without worsening the
corresponding fit to other data, they did not consider the $R_{\rm h}=ct$
cosmology. Our results here show that, insofar as the AP test by itself
is concerned, this model (and coincidentally the STL model) does in fact
yield a better fit to the BOSS measurements than does the standard model.

A comparison between these results and those reported in ref.~\cite{17}
highlights the dramatic improvement in the measurement of the BAO peak position
that has led to the important conclusions drawn in this paper. Though the measurements
of $y(z)$ based on the galaxy two-point correlation function used in that earlier
work were good enough to rule out all but the concordance and STL models
(the $R_{\rm h}=ct$ universe was not included in that comparison, but
its $y(z)$ function is identical to that of STL), the errors arising
from the contamination due to internal redshift distortions were still too large
for us to discriminate between these two cosmologies. As we can see from
figure~1, however, the significant improvement in the precision with which the
BAO peak position is measured now\cite{Ala16,27} disfavors the standard model.

\subsection{The Acoustic Scale}
\label{.acousticscale}
The surprising and emphatic result summarized in figure~1 now compels us
to examine one of $\Lambda$CDM's most impressive successes---the interpretation 
of the CMB and BAO lengths as arising from a single, consistent acoustic scale---and
whether this identification survives into the $R_{\rm h}=ct$ framework. 
An acoustic angular size, $\theta_s=(0.596724 \pm 0.00038)^\circ$, has been 
measured in both the temperature and polarization power spectrum, most recently
with {\it Planck}\cite{Ade14}. This `standard ruler' is believed to be 
responsible not only for the multi-peak structure in the CMB power
spectrum, but also for leaving its indelible imprint on the large-scale structure we
see today. In many ways, the optimization of the parameters in $\Lambda$CDM
relies critically on the correct interpretation of this `sound horizon.'

One may question some of the assumptions made in calculating the acoustic
scale in $\Lambda$CDM, including its estimation from the comoving distance
$r_s$ traveled by a sound wave rather than from an actual solution to the
geodesic equation in an expanding medium, but the fact that 
$r_s^{\rm CMB}/r_s^{\rm BAO}\sim 1$ is evidence in support of the
standard model. Let us now see what happens in the $R_{\rm h}=ct$ universe.
The measured value of $r_s^{\rm BAO}$ at $z=2.34$ is $\sim 144.8\pm8$ Mpc, assuming the 
{\it Planck} Hubble constant $H_0=67.8$ km s$^{-1}$ Mpc$^{-1}$ (ref.~\cite{Ade14}). 
We have recently initiated an analysis of the CMB power spectrum and its
angular correlation function\cite{MSL2016} in the context of $R_{\rm h}=ct$,
which has yielded an optimized value of $\approx 3.2/2\pi$ for the ratio 
$u_{\rm min}/2\pi$ of the angular-diameter distance $d_A(t_{\rm dec})$ over the
maximum fluctuation size $\lambda_{\rm max}$ at decoupling. In $R_{\rm h}=ct$,
$\lambda_{\rm max}=2\pi c/H(t_{\rm dec})$, where $H(t_{\rm dec})$ is the
Hubble constant at the last scattering surface. Therefore, the comoving
CMB acoustic scale in this model is given by the simple expression 
$r_s^{\rm CMB}=\theta_s(c/H_0)u_{\rm min}$ (remembering that $H(t)=1/t$
and $(1+z)=t_0/t$ in this cosmology). That is, $r_s^{\rm CMB}\approx 148$
Mpc, which is fully consistent with the BAO value $144.8\pm8$ Mpc
measured at $z=2.34$.

If the results of the AP test are reliable, and $R_{\rm h}=ct$ is indeed 
the correct cosmology, it is therefore likely that the CMB and BAO length scales 
have a common origin, just as they apparently do in $\Lambda$CDM as well.
It is therefore highly desirable to carry out additional high-precision 
measurements of the BAO scale using a broader redshift coverage than is 
currently available. A quick inspection of figure~1 suggests that to 
distinguish between the various models, the AP test is best suited to 
redshifts $z\gtrsim 2$---the higher the better.

\section{Conclusions}
The results of our analysis in this paper show that, if the measurements of
$d_A(z)$ and $H(z)$ derived from the BAO peak anisotropic distributions
are correct, the concordance $\Lambda$CDM model, optimized to fit SNIa and
CMB data, does not pass the AP test. In light of the AP results, 
one may begin to question its status as a true `concordance' model. 
Instead, the AP test using these model-independent
data favors the $R_{\rm h}=ct$ universe, which has thus far also been
favored by model selection tools in other one-on-one comparative tests
with $\Lambda$CDM\cite{31,32,33,34,35,36,37,38,41,42}.
The consequences of this important result are being explored elsewhere,
including the growing possibility that inflation may have been unnecessary
to resolve any perceived `horizon problem' and therefore may have simply
never happened\cite{33}.

\section*{Acknowledgments}
We are grateful to Chao-Jun Feng for pointing out an error in an earlier version 
of this manuscript, and to the anonymous referee for his thoughtful and helpful 
review that has led to an improvement in the presentation of our results. We are 
particularly grateful to Andreu Font-Ribera for his helpful private comments on 
the interpretation of his paper\cite{21}. MLC was supported by grant AYA2015-66506-P 
of the Spanish Ministry of Economy and Competitiveness (MINECO). F.M. is grateful 
to Amherst College for its support through a John Woodruff Simpson Lectureship, 
and to Purple Mountain Observatory in Nanjing, China, for its hospitality while 
part of this work was being carried out. This work was partially supported by 
grant 2012T1J0011 from The Chinese Academy of Sciences Visiting Professorships 
for Senior International Scientists, and grant GDJ20120491013 from the Chinese 
State Administration of Foreign Experts Affairs.

\end{document}